\documentclass[12pt]{article}

\begin{document}

\title{\bf Kinematic Self-Similar Plane Symmetric Solutions}

\author{M. Sharif \thanks{msharif@math.pu.edu.pk} and Sehar Aziz
\thanks{sehar$\_$aziz@yahoo.com}\\
Department of Mathematics, University of the Punjab,\\
Quaid-e-Azam Campus, Lahore-54590, Pakistan.}

\date{}

\maketitle

\begin{abstract}
This paper is devoted to classify the most general plane symmetric
spacetimes according to kinematic self-similar perfect fluid and
dust solutions. We provide a classification of the kinematic
self-similarity of the first, second, zeroth and infinite kinds
with different equations of state, where the self-similar vector
is not only tilted but also orthogonal and parallel to the fluid
flow. This scheme of classification yields twenty four plane
symmetric kinematic self-similar solutions. Some of these
solutions turn out to be vacuum. These solutions can be matched
with the already classified plane symmetric solutions under
particular coordinate transformations. As a result, these reduce
to sixteen independent plane symmetric kinematic self-similar
solutions.
\end{abstract}


\section{Introduction}

Einstein field equations (EFEs) are highly non-linear, second order
coupled partial differential equations and hence could not be solved
unless certain symmetry assumptions are taken on the spacetime
metric. There has been a recent literature [1-7, and references
therein] which shows a significant interest in the study of various
symmetries. Self-similarity leads to ordinary differential equations
(ODEs) and their mathematical treatment is relatively simple.
Invariance of the field equations under a scale transformation
indicates that there exist scale invariant solutions to the EFEs.
These solutions are known as self-similar solutions. Although
self-similar solutions are only special solutions, they often play
an important role in cosmological situations and gravitational
collapse.

There exist several preferred geometric structures in self-similar
models and a number of natural approaches may be used in studying
them. The three most common ones are the co-moving, homothetic and
Schwarzschild approaches. Each of the approaches has its individual
physical interpretational advantages and they are all complementary.
In the co-moving approach, pioneered by Cahill and Taub [8], the
coordinates are adopted to the fluid 4-velocity vector. This
probably affords the best physical insight and is the most
convenient one. In General Relativity (GR), a self-similarity
defined by the existence of a homothetic vector (HV) field is called
self-similarity of the first kind (or homothety or continuous self
similarity). There exists a natural generalization of homothety
called kinematic self-similarity, which is defined by the existence
of a kinematic self-similar (KSS) vector field.

Cahill and Taub [8] gave the concept of self-similarity
corresponding to Newtonian self-similarity of the homothetic class.
Carter and Henriksen [9,10] defined self-similarity of the second,
zeroth and infinite kinds. The only compatible barotropic equation
of state with self-similarity of the first kind is $p=k\rho.$ The
classification of the self-similar perfect fluid solutions of the
first kind in the dust case ($k=0$) has been provided by Carr [1].
The case $0<k<1$ has been studied by Carr and Coley [2]. Coley [11]
has shown that the FRW solution is the only spherically symmetric
homothetic perfect fluid solution in the parallel case. McIntosh
[12] has discussed that a stiff fluid ($k=1$) is the only compatible
perfect fluid with the homothety in the orthogonal case. Benoit and
Coley [13] have studied analytic spherically symmetric solutions of
the EFEs coupled with a perfect fluid and admitting a KSS vector of
the first, second and zeroth kinds. Sintes et al. [14] have
considered spherically, plane and hyperbolic symmetric spacetimes
which admit a KSS vector of the infinite kind with perfect fluid.
Carr et al. [15,16] have explored the KSS vector associated with the
critical behavior observed in the gravitational collapse of
spherically symmetric perfect fluid with equation of state
$p=k\rho$. Further, they have investigated solution space of
self-similar spherically symmetric perfect fluid models and physical
aspects of these solutions. Coley and Goliath [17] have investigated
self-similar spherically symmetric cosmological models with a
perfect fluid and a scalar field with an exponential potential.

The assumption of self-similarity is very powerful in finding
analytical solutions. The group $G_3$ contains two special cases of
physical interest, i.e., spherical and plane symmetry. Most of the
literature is available on spherical symmetric spacetimes. Maeda et
al. [3,4] have studied the KSS vector of the second, zeroth and
infinite kinds in the tilted, parallel and orthogonal cases. The
same authors [5] have also discussed the classification of the
spherically symmetric KSS perfect fluid and dust solutions. This
analysis has provided some interesting solutions. Recently, Sharif
and Sehar [7] have investigated the KSS solutions for the
cylindrically symmetric spacetimes. The analysis has been
extensively given for the perfect fluid and dust cases with tilted,
parallel and orthogonal vectors by using different equations of
state. Some interesting consequences have been developed. The same
authors have also studied the properties of such solutions for
spherically symmetric [18], cylindrically symmetric [19] and plane
symmetric spacetimes [20].

In a recent paper, Sharif and Sehar [21] have explored the KSS
solutions for the plane symmetric spacetimes under certain
assumption. The investigation is incomplete due to this restriction
on plane symmetric spacetimes. In this paper, we drop this
restriction and deal with the most general plane symmetric
spacetimes. This analysis provides many more interesting
self-similar solutions. The paper has been organised as follows. In
section 2, we briefly review KSS vector of different kinds
corresponding to the plane symmetric spacetimes. Sections 3 and 4
are devoted to the titled perfect fluid and dust solutions
respectively. The orthogonal perfect fluid and dust solutions are
investigated in section 5. Sections 6 and 7 are used to explore the
parallel perfect fluid and dust cases respectively. In the last
section, we present summary of the results and their discussion.

\section{Plane Symmetric Spacetime and Kinematic Self-Similarity}

A plane symmetric spacetime is a Lorentzian manifold possessing a
physical stress-energy tensor. This admits $SO(2)(\times\Re^2$ as
the minimal isometry group in such a way that the group orbits are
spacelike surfaces of constant curvature. The most general plane
symmetric metric is given in the form [22]
\begin{equation}
ds^2=e^{2\nu(t,x)}dt^2-e^{2\mu(t,x)}dx^2-e^{2\lambda(t,x)}(dy^2+ dz^2),
\end{equation}
where $\nu$, $\mu$ and $\lambda$ are arbitrary functions of $t$
and $x$. The energy-momentum tensor for a perfect fluid can be written as
\begin{equation}
T_{ab}=[\rho(t,x)+p(t,x)]u_{a}u_{b}-p(t,x)g_{ab},\quad
(a,b=0,1,2,3),
\end{equation}
where $\rho$ is the density, $p$ is the pressure and $u_{a}$ is the
four-velocity of the fluid element. In the co-moving coordinate
system, the four-velocity can be written as
$u_{a}=(e^{\nu(t,x)},0,0,0)$. Using Eqs.(1) and (2), we can write
the EFEs as
\begin{eqnarray}
\kappa\rho&=&e^{-2\mu}(2\lambda_x\mu_x-3{\lambda_{x}}^{2}-2\lambda_{xx})
+e^{-2\nu}(2\lambda_t\mu_t+{\lambda_{t}}^{2}),\\
0&=&\lambda_{tx}-\lambda_{t}\nu_{x}+\lambda_{t}\lambda_{x}-\lambda_x\mu_t,\\
\kappa p &=& e^{-2\mu}({\lambda_{x}}^{2}+2\lambda_{x}\nu_{x})
-e^{-2\nu}(2\lambda_{tt}-2\lambda_{t}\nu_{t}+3{\lambda_{t}}^{2}),\\
\kappa p &=& e^{-2\mu}(\nu_{xx}+{\nu_{x}}^{2}+\nu_{x}\lambda_{x}+{\lambda_{x}}^{2}
+\lambda_{xx}-\lambda_{x}\mu_{x}-\nu_{x}\mu_{x})\nonumber\\
&-&e^{-2\nu}(\lambda_{tt}-\lambda_{t}\nu_{t}+{\lambda_{t}}^{2}+\mu_{tt}
+{\mu_{t}}^2+\lambda_{t}\mu_{t}-\nu_{t}\mu_{t}).
\end{eqnarray}
It follows from the conservation of energy-momentum tensor,
${T^{ab}}_{;b}=0$, that
\begin{equation}
\mu_{t}=-\frac{\rho_{t}}{(\rho+p)}-2\lambda_{t},
\end{equation}
and
\begin{equation}
\nu_{x}=-\frac{p_{x}}{(\rho+p)}.
\end{equation}
For a plane symmetric spacetime, the general form of a vector field
$\xi$ can be written as
\begin{equation}
\xi^{a}\frac{\partial}{\partial
x^a}=h_{1}(t,x)\frac{\partial}{\partial
t}+h_{2}(t,x)\frac{\partial}{\partial x},
\end{equation}
where $h_{1}$ and $h_{2}$ are arbitrary functions. The vector field $\xi$
can have three cases, i.e., parallel, orthogonal and tilted. They are
distinguished by the relation between the generator and a timelike vector
field, which is identified as the fluid flow, if it exists.  When $\xi$ is
parallel to the fluid flow, $h_{2}=0$ and when $\xi$ is orthogonal to the
fluid flow $h_{1}=0$. When both $h_{1}$ and $h_{2}$ are non-zero, $\xi$ is
tilted to the fluid flow. The tilted case is the most general among them.

We define a KSS vector $\xi$ such that
\begin{eqnarray}
\pounds_{\xi}h_{ab}&=& 2\delta h_{ab},\\
\pounds_{\xi}u_{a}&=& \alpha u_{a},
\end{eqnarray}
where $h_{ab}=g_{ab}-u_au_b$ is the projection tensor, $\alpha$ and
$\delta$ are constants. The ratio, $\alpha/\delta$, is called the
similarity index which gives rise to the following two cases:
\begin{eqnarray*}
1. \quad \delta\neq0;\quad 2. \quad \delta=0.
\end{eqnarray*}
{\bf Case 1:} When $\delta\neq0$, we can choose it as unity and the
KSS vector for the titled case can take the following form
\begin{equation}
\xi^{a}\frac{\partial}{\partial x^a}=(\alpha
t+\beta)\frac{\partial}{\partial t}+x\frac{\partial}{\partial x}.
\end{equation}
In this case, the similarity index, $\alpha/\delta$, further implies
the following three possibilities
\begin{eqnarray*}
&(i)&\quad\delta\neq0,\quad\alpha=1\quad
(\beta~can~be~taken~to~be~zero),\\
&(ii)&\quad\delta\neq0,\quad\alpha=0\quad
(\beta~can~be~taken~to~be~unity),\\
&(iii)&\quad\delta\neq0,\quad\alpha\neq0,1\quad
(\beta~can~be~taken~to~be~zero).
\end{eqnarray*}
The first case 1(i) is referred to the self-similarity of the {\it
first kind}. In this case, $\xi$ is a homothetic vector and the
self-similar variable $\xi$ turns out to be $x/t$. For the second
case 1(ii), it is called the self-similarity of the {\it zeroth
kind} and the self-similar variable becomes $\xi=x e^{-t}$. The last
case 1(iii) is known as the self-similarity of the {\it second kind}
and the self-similar variable turns out to be
\begin{eqnarray*}
\xi=\frac{x}{(\alpha t)^\frac{1}{\alpha}}.
\end{eqnarray*}
For the case (1), the metric functions take the following form
\begin{equation}
\nu=\nu(\xi),\quad \mu=\mu(\xi),\quad e^{\lambda}=xe^{\lambda(\xi)}.
\end{equation}
{\bf Case 2:} In the second case (2), when $\delta=0$ and
$\alpha\neq0$ ($\alpha$ can be unity and $\beta$ can be re-scaled to
zero), the self-similarity is referred to the {\it infinite kind}.
Here, the KSS vector $\xi$ becomes
\begin{equation}
\xi^{a}\frac{\partial}{\partial x^a}=t\frac{\partial}{\partial
t}+x\frac{\partial}{\partial x}
\end{equation}
and the self-similar variable takes the form $\xi=x/t$.
Consequently, the metric functions will become
\begin{equation}
\nu=\nu(\xi),\quad \mu=-\ln{(x)}+\mu(\xi),\quad\lambda=\lambda(\xi).
\end{equation}
It is mentioned here that for $\delta=0=\alpha$, the KSS vector
$\xi$ reduces to KV.

When the KSS vector $\xi$ is parallel to the fluid flow, we obtain
\begin{equation}
\xi^{a}\frac{\partial}{\partial x^a}= f(t)\frac{\partial}{\partial
t},
\end{equation}
where $f(t)$ is an arbitrary function. It is worth mentioning point
here that we obtained [21] contradictory results in the first,
second and zeroth kinds while for the infinite kind the self-similar
variable was $x$. As a result, there was no solution when $\xi$ was
parallel to the fluid flow in the first, second and zeroth kinds
except for the infinite kind. However, this analysis of the most
general plane symmetric spacetimes yields self-similar variable $x$
in each kind and hence we can expect solution. The metric functions
for the first, second, zeroth and infinite kinds, respectively, will
be
\begin{eqnarray}
\nu&=&\nu(x),\quad
\mu=\ln{(t)}+\mu(x),\quad\lambda=\ln{(t)}+\lambda(x),\nonumber\\
\nu&=&(\alpha-1)\ln{(t)}+\nu(x),\quad
\mu=\ln{(t)}+\mu(x),\quad\lambda=\ln{(t)}+\lambda(x),\nonumber\\
\nu&=&-\ln{(t)}+\nu(x),\quad \mu=\ln{(t)}+\mu(x),\quad
\lambda=\ln{(t)}+\lambda(x),\nonumber\\
\nu&=&\nu(x),\quad \mu=\mu(x),\quad\lambda=\lambda(x).
\end{eqnarray}

If the KSS vector $\xi$ is orthogonal to the fluid flow, it follows
that
\begin{equation}
\xi^{a}\frac{\partial}{\partial x^a}= g(x)\frac{\partial}{\partial
x},
\end{equation}
where $g(x)$ is an arbitrary function and the self-similar variable
is $t$. The corresponding metric functions for the first, second,
zeroth and infinite kinds, respectively, will take the following
form
\begin{eqnarray}
\nu&=&\ln{(x)}+\nu(t),\quad
\mu=\mu(t),\quad\lambda=\ln{(x)}+\lambda(t),\nonumber\\
\nu&=&\alpha\ln{(x)}+\nu(t),\quad
\mu=\mu(t),\quad\lambda=\ln{(x)}+\lambda(t),\nonumber\\
\nu&=&\nu(t),\quad
\mu=\mu(t),\quad\lambda=\ln{(x)}+\lambda(t),\nonumber\\
\nu&=&\ln{(x)}+\nu(t),\quad
\mu=-\ln{(x)}+\mu(t),\quad\lambda=\lambda(t).
\end{eqnarray}
The following two types of polytropic equations of state (EOS) will
be assumed. The first equation of state, denoted by EOS(1), is
\begin{eqnarray*}
p=k\rho^{\gamma},
\end{eqnarray*}
where $k$ and $\gamma$ are constants. Another EOS is the following
[17]
\begin{eqnarray*}
p=kn^{\gamma},\quad \rho=m_{b}n+\frac{p}{\gamma-1},
\end{eqnarray*}
where $m_{b}$ is a constant which corresponds to the baryon mass,
and $n(t,r)$ corresponds to baryon number density. We call this
equation as the second equation of state EOS(2). Notice that we take
$k\neq0$ and $\gamma\neq0,1$ for EOS(1) and EOS(2). EOS(3) is given
by
\begin{eqnarray*}
p=k\rho,\quad -1\leq k \leq 1,\quad k\neq0.
\end{eqnarray*}

\section{Tilted Perfect Fluid Case}

\subsection{Self-Similarity of the First Kind}

It follows from the EFEs that the energy density $\rho$ and pressure
$p$ must take the following form
\begin{eqnarray}
\kappa\rho &=& \frac{1}{x^2}\rho(\xi),\\
\kappa p &=& \frac{1}{x^2}p(\xi),
\end{eqnarray}
where the self-similar variable is $\xi=x/t$. When the EFEs and the
equations of motion for the matter field are satisfied, it yields a
set of ODEs and hence Eqs.(3)-(8) reduce to
\begin{eqnarray}
\dot{\rho} &=&-(\dot{\mu}+2\dot{\lambda})(\rho+p),\\
2p-\dot{p} &=& \dot{\nu}(\rho+p),\\
e^{2\mu}\rho &=& 2\dot{\mu}+2\dot{\lambda}\dot{\mu}-4\dot{\lambda}
-3{\dot{\lambda}}^{2}-2\ddot{\lambda}-1,\\
0&=& 2\dot{\lambda}\dot{\mu}+{\dot{\lambda}}^{2},\\
0&=&\ddot{\lambda}+{\dot{\lambda}}^2+\dot{\lambda}-\dot{\mu}
-\dot{\lambda}\dot{\mu}-\dot{\lambda}\dot{\nu},\\
e^{2\mu}p&=&1+2\dot{\lambda}+{\dot{\lambda}}^2
+2\dot{\nu}+2\dot{\lambda}\dot{\nu},\\
0&=& 2\dot{\lambda}\dot{\nu} -2\ddot{\lambda}-3{\dot{\lambda}}^2
-2\dot{\lambda},\\
e^{2\mu}p &=& \ddot{\lambda}+{\dot{\lambda}}^2 +\dot{\lambda}
+\dot{\lambda}\dot{\nu}+\ddot{\nu}+{\dot{\nu}}^2
-\dot{\mu}-\dot{\lambda}\dot{\mu}-\dot{\nu}\dot{\mu},\\
0&=&-\ddot{\lambda}-{\dot{\lambda}}^2 -\dot{\lambda}-\ddot{\mu}
-{\dot{\mu}}^2-\dot{\mu}+\dot{\lambda}\dot{\nu}
-\dot{\lambda}\dot{\mu}+\dot{\mu}\dot{\nu}.
\end{eqnarray}
Here dot means derivative with respect to $ln(\xi)$.

\subsubsection{Equation of State (3)}

If a perfect fluid satisfies EOS(3), Eqs.(20) and (21) yield that
\begin{equation}
p=k\rho. \quad [Case~I]
\end{equation}
From Eq.(25), we have two possibilities either $\dot{\lambda}=0$ or
$\dot{\lambda}=-2\dot{\mu}$. For the first possibility, we obtain
the following vacuum solution
\begin{eqnarray}
\nu&=&\ln{(c_0\xi^{(1\mp\sqrt{2})})},\quad \mu=c_1, \quad \lambda=c_2,\nonumber\\
\rho&=&constant,\quad k=-3\pm\sqrt{2}.
\end{eqnarray}
The corresponding metric is
\begin{equation}
ds^2=(\frac{x}{t})^{(2\mp2\sqrt{2})}dt^2-dx^2-x^2(dy^2+ dz^2).
\end{equation}
The second possibility leads to contradiction.

\subsection{Self-Similarity of the Second Kind}

It follows from the EFEs that the energy density $\rho$ and pressure
$p$ can be written as
\begin{eqnarray}
\kappa\rho &=& \frac{1}{x^2}[\rho_1(\xi)+\frac{x^2}{t^2}\rho_2(\xi)],\\
\kappa p &=& \frac{1}{x^2}[p_1(\xi)+\frac{x^2}{t^2}p_2(\xi)],
\end{eqnarray}
where the self-similar variable is $\xi=x/(\alpha
t)^\frac{1}{\alpha}$. When the EFEs and the equations of motion for
the matter field are satisfied for $O[(\frac{x}{t})^0]$ and
$O[(\frac{x}{t})^2]$ terms separately, we obtain the following ODEs
\begin{eqnarray}
\dot{\rho_1} &=&-(\dot{\mu}+2\dot{\lambda})(\rho_1+p_1),\\
\dot{\rho_2}+2\alpha\rho_2 &=&-(\dot{\mu}+2\dot{\lambda})(\rho_2+p_2),\\
-\dot{p_1}+2p_1 &=& \dot{\nu}(\rho_1+p_1),\\
-\dot{p_2} &=&\dot{\nu}(\rho_2+p_2) ,\\
e^{2\mu}\rho_1 &=& 2\dot{\mu}+2\dot{\mu}\dot{\lambda}-4\dot{\lambda}
-3{\dot{\lambda}}^{2}-2\ddot{\lambda}-1,\\
\alpha^2e^{2\nu}\rho_2 &=& 2\dot{\mu}\dot{\lambda}+{\dot{\lambda}}^{2},\\
0&=&\ddot{\lambda}+{\dot{\lambda}}^2+\dot{\lambda}
-\dot{\mu}-\dot{\lambda}\dot{\nu}-\dot{\lambda}\dot{\mu},\\
e^{2\mu}p_1&=&1+2\dot{\lambda}+{\dot{\lambda}}^2+2\dot{\nu}
+2\dot{\lambda}\dot{\nu},\\
\alpha^2e^{2\nu}p_2 &=& -2\ddot{\lambda}-3{\dot{\lambda}}^2
-2\alpha\dot{\lambda}+2\dot{\lambda}\dot{\nu},\\
e^{2\mu}p_1&=&\ddot{\lambda}+{\dot{\lambda}}^2
+\dot{\lambda}+\dot{\lambda}\dot{\nu}+\ddot{\nu}+{\dot{\nu}}^2
-\dot{\mu}-\dot{\mu}\dot{\lambda}-\dot{\nu}\dot{\mu},\\
\alpha^2e^{2\nu}p_2&=&-\ddot{\lambda}-{\dot{\lambda}}^2
-\alpha\dot{\lambda}-\ddot{\mu}-{\dot{\mu}}^2
-\alpha\dot{\mu}+\dot{\lambda}\dot{\nu}+\dot{\mu}\dot{\nu}
-\dot{\lambda}\dot{\mu}.
\end{eqnarray}

\subsubsection{Equations of State (1) and (2)}

When a perfect fluid satisfies EOS(1) for $k\neq0$ and $\gamma\neq0,1$, Eqs.(34) and
(35) become
\begin{equation}
\alpha=\gamma,\quad p_1=0=\rho_2,\quad p_2= \frac{k}{(8\pi
G)^{(\gamma-1)}\gamma^2}\xi^{-2\gamma}{\rho_1}^\gamma,
\quad[Case~I]
\end{equation}
or
\begin{equation}
\alpha=\frac{1}{\gamma},\quad p_2=0=\rho_1,\quad
p_1=\frac{k}{(8\pi
G)^{(\gamma-1)}\gamma^{2\gamma}}\xi^{2}{\rho_2}^\gamma.\quad[Case~II]
\end{equation}
For a perfect fluid with EOS(2) and $k\neq0,~\gamma\neq0,1$, it follows from Eqs.(34)
and (35) that
\begin{equation}
\alpha=\gamma,\quad p_1=0,\quad p_2=\frac{k}{{m_b}^{\gamma}(8\pi
G)^{(\gamma-1)}\gamma^2}\xi^{-2\gamma}{\rho_1}^\gamma=(\gamma-1)\rho_2,
\quad [Case~III]
\end{equation}
or
\begin{equation}
\alpha=\frac{1}{\gamma}, \quad p_2=0, \quad
p_1=\frac{k}{{m_b}^{\gamma}(8\pi
G)^{(\gamma-1)}\gamma^{2\gamma}}\xi^{2}{\rho_2}^\gamma=(\gamma-1)\rho_1.
\quad [Case~IV]
\end{equation}
In the \textbf{Case $I$}, Eq.(37) gives rise to two possibilities,
i.e., either $\dot{\mu}=-2\dot{\lambda}$ or $p_2=0$. For the first
possibility we meet a contradiction. In the second option, we obtain
the following vacuum solution
\begin{eqnarray}
\nu&=&c_1,\quad \mu=\frac{1}{2}\ln{\xi}+c_2,
\quad  \lambda=-\ln{\xi}+c_3,\nonumber\\
\rho_1&=&0=p_1,\quad \rho_2=0=p_2,\quad \alpha=\frac{3}{2}.
\end{eqnarray}
The corresponding metric is
\begin{equation}
ds^2=dt^2-\frac{2^{2/3}x}{(3t)^{2/3}}dx^2
-(\frac{3t}{2})^{4/3}(dy^2+ dz^2).
\end{equation}
For the \textbf{Case $II$}, Eq.(36) shows that either
$\dot{\mu}=-2\dot{\lambda}$ or $p_1=0$. The first possibility leads
to contradiction and the second possibility yields the same solution
as given by Eq.(51).

In the \textbf{Case $III$}, Eq.(38) implies that either $\rho_1=0$
or $\dot{\nu}=0$. For the first option, Eq.(41) implies that either
$\dot{\lambda}=0$ or $\dot{\lambda}=-2\dot{\mu}$. The case when
$\dot{\lambda}=0$ gives a contradiction and the option
$\dot{\lambda}=-2\dot{\mu}$ implies the same solution as given by
Eq.(51). The second case $\dot{\nu}=0$ and the \textbf{Case $IV$}
also lead to the same solution as Eq.(51).

\subsubsection{Equation of State (3)}

For a perfect fluid satisfying EOS(3), Eqs.(34) and (35) yield that
\begin{equation}
p_1=k\rho_1, \quad p_2=k\rho_2. \quad [Case~V]
\end{equation}
This implies two options either $k=-1$ or $k\neq-1$. For $k=-1$,
Eqs.(36)-(46) lead to the same solution as for EOS(1) and EOS(2)
given by Eq.(51). For $k\neq-1$, the case $\rho_1\neq0,~\rho_2\neq0$
leads to a contradiction. The case, when $\rho_1=0$ and $\rho_2$ is
arbitrary, implies that
\begin{eqnarray}
\nu&=&c_1,\quad \mu=\frac{3}{2}\ln{\xi}+c_2,
\quad  \lambda=-\ln{\xi}+c_3,\nonumber\\
\rho_1&=&0=p_1,\quad \rho_2=constant=p_2, \quad \alpha=\frac{1}{2}.
\end{eqnarray}
The resulting plane symmetric metric becomes
\begin{equation}
ds^2=dt^2-\frac{64x^3}{t^6}dx^2-\frac{t^4}{16}(dy^2+ dz^2).
\end{equation}
For the case when $\rho_2=0$ and $\rho_1$ is arbitrary, Eq.(41)
implies that either $\dot{\lambda}=0$ or
$\dot{\lambda}=-2\dot{\mu}$. For the first possibility, it follows
that
\begin{eqnarray}
\nu&=&\frac{2k}{k+1}\ln{\xi}+c_1,\quad \mu=c_2,\quad \lambda=c_3,
\quad p_1=constant, \nonumber\\\rho_2&=&0=p_2,\quad k=-3\pm2\sqrt{2}
\end{eqnarray}
and hence the plane symmetric spacetime will take the following form
\begin{equation}
ds^2=(\frac{x}{(\alpha
t)^{1/\alpha}})^\frac{4k}{k+1}dt^2-dx^2-x^2(dy^2+ dz^2).
\end{equation}
For the second possibility, Eqs.(42) and (44) further imply two
possibilities either $\dot{\mu}=0$ or $\alpha=\frac{3}{2}$. When
$\dot{\mu}=0$ we obtain the same solution as Eq.(56). For
$\alpha=\frac{3}{2}$, we can solve the system of equations by
assuming either $\dot{\mu}=0$ or $\dot{\nu}=0$. Assuming
$\dot{\mu}=0$, we obtain the same solution as given by Eq.(56). If
we take $\dot{\nu}=0$, we have a contradiction.

\subsection{Self-Similarity of the Zeroth Kind}

For this case, the EFEs show that the quantities $\rho$ and $p$ must
be of the form
\begin{eqnarray}
\kappa\rho &=& \frac{1}{x^2}[\rho_1(\xi)+x^2\rho_2(\xi)],\\
\kappa p &=& \frac{1}{x^2}[p_1(\xi)+x^2p_2(\xi)],
\end{eqnarray}
where the self-similar variable is $\xi=\frac{x}{e^{t}}$. Assuming
that the EFEs and the equations of motion for the matter field are
satisfied for $O[(x)^0]$ and $O[(x)^2]$ terms separately, it follows
that
\begin{eqnarray}
\dot{\rho_1} &=&-(\dot{\mu}+2\dot{\lambda})(\rho_1+p_1),\\
\dot{\rho_2} &=&-(\dot{\mu}+2\dot{\lambda})(\rho_2+p_2),\\
-\dot{p_1}+2p_1 &=& \dot{\nu}(\rho_1+p_1),\\
-\dot{p_2} &=&\dot{\nu}(\rho_2+p_2) ,\\
e^{2\mu}\rho_1 &=& 2\dot{\mu}-4\dot{\lambda}-3{\dot{\lambda}}^{2}
-2\ddot{\lambda}+2\dot{\lambda}\dot{\mu}-1,\\
e^{2\nu}\rho_2 &=& 2\dot{\lambda}\dot{\mu}+{\dot{\lambda}}^{2},\\
0&=&\ddot{\lambda}+{\dot{\lambda}}^2+\dot{\lambda}-\dot{\mu}
-\dot{\lambda}\dot{\mu}-\dot{\lambda}\dot{\nu},\\
e^{2\mu}p_1&=&1+2\dot{\lambda}+{\dot{\lambda}}^2
+2\dot{\nu}+2\dot{\lambda}\dot{\nu},\\
e^{2\nu}p_2&=&2\dot{\lambda}\dot{\nu}
-2\ddot{\lambda}-3{\dot{\lambda}}^2,\\
e^{2\mu}p_1&=&\ddot{\lambda}+{\dot{\lambda}}^2
+\dot{\lambda}+\dot{\lambda}\dot{\nu}
+\ddot{\nu}+{\dot{\nu}}^2-\dot{\mu}
-\dot{\lambda}\dot{\mu}-\dot{\mu}\dot{\nu},\\
e^{2\nu}p_2&=&-\ddot{\lambda}-{\dot{\lambda}}^2
+\dot{\lambda}\dot{\nu}
-\dot{\lambda}\dot{\mu}+\dot{\mu}\dot{\nu}
-{\dot{\mu}}^2-\ddot{\mu}.
\end{eqnarray}

\subsubsection{EOS(1) and EOS(2)}

Here Eqs.(58) and (59) imply that
\begin{equation}
p_1=0=\rho_1,\quad p_2= \frac{k}{(8\pi
G)^{(\gamma-1)}}{\rho_2}^\gamma, \quad[Case~I]
\end{equation}
For EOS(2), it turns out that
\begin{equation}
p_1=0=\rho_1,\quad p_2=\frac{k}{{m_b}^{\gamma}(8\pi
G)^{(\gamma-1)}}[{\rho_2}-\frac{p_2}{(\gamma-1)}]^\gamma. \quad
[Case~II]
\end{equation}
In both cases, we get the same set of equations which on solving
yield the following solution for both EOS(1) and EOS(2)
\begin{eqnarray}
\nu&=&c_1,\quad \mu=-\ln{\xi}+\ln{(\xi^3-c_3)}+c_2,\quad
\lambda=-\ln{\xi}+c_4,\nonumber\\
\rho_1&=&0=p_1,\quad
\rho_2=-\frac{3(\xi^3+c_3)}{e^{2c_1}(\xi^3-c_3)}, \quad
p_2=constant.
\end{eqnarray}
The corresponding metric is
\begin{equation}
ds^2=dt^2-(\frac{x^3-c_3e^{3t}}{xe^{2t}})^2dx^2-e^{2t}(dy^2+ dz^2).
\end{equation}

\subsubsection{EOS(3)}

Here it follows from Eqs.(58) and (59) that
\begin{equation}
p_1=k\rho_1,\quad  p_2=k\rho_2.
\end{equation}
Proceeding in a similar fashion as in the case of self-similarity
of the second kind with EOS(3), we obtain, for $k=-1$, the
following solution
\begin{eqnarray}
\nu&=&c_1,\quad \mu=-\ln{\xi}+c_2,
\quad \lambda=-\ln{\xi}+c_3,\nonumber\\
\rho_1&=&0=p_1,\quad \rho_2=constant=-p_2.
\end{eqnarray}
The corresponding plane symmetric metric is
\begin{equation}
ds^2=dt^2-\frac{e^{2t}}{x^2}dx^2-e^{2t}(dy^2+ dz^2).
\end{equation}
The case $k\neq-1$ again leads to three options either
$\rho_1\neq0\neq\rho_2,~\rho_1=0$ or $\rho_2=0$. The first case
gives a contradiction. For the second option, we obtain the
following solution
\begin{eqnarray}
\nu&=&c_1,\quad \mu=2\ln{\xi}+c_2,
\quad  \lambda=-\ln{\xi}+c_3,\nonumber\\
\rho_1&=&0=p_1,\quad \rho_2=constant=p_2.
\end{eqnarray}
The plane symmetric metric for this solution becomes
\begin{equation}
ds^2=dt^2-x^4e^{-4t}dx^2-e^{2t}(dy^2+ dz^2).
\end{equation}
The case, when $\rho_2=0$, Eq.(65) yields two possibilities either
$\dot{\lambda}=0$ or $\dot{\lambda}=-2\dot{\mu}$. For both
possibilities, we obtain the same solution as in the case of the
second kind with EOS(3) given by Eq.(56) ($\alpha=0$). The
corresponding metric is
\begin{equation}
ds^2=(xe^{-t})^\frac{4k}{k+1}dt^2-dx^2-e^{2t} (dy^2+ dz^2).
\end{equation}

\subsection{Self-Similarity of the Infinite Kind}

In this case, the EFEs indicate that the quantities $\rho$ and $p$
must be of the following form
\begin{eqnarray}
\kappa\rho &=& \rho_1(\xi)+\frac{1}{t^2}\rho_2(\xi),\\
\kappa p &=& p_1(\xi)+\frac{1}{t^2}p_2(\xi),
\end{eqnarray}
where $\xi=\frac{x}{t}$.  The requirement that the EFEs and the
equations of motion for the matter field are satisfied for
$O[(t)^0]$ and $O[(t)^{-2}]$ terms separately leads to the following
equations
\begin{eqnarray}
\dot{\rho_1} &=&-(\dot{\mu}+2\dot{\lambda})(\rho_1+p_1),\\
\dot{\rho_2}+2\rho_2 &=&-(\dot{\mu}+2\dot{\lambda})(\rho_2+p_2),\\
-\dot{p_1} &=&\dot{\nu}(\rho_1+p_1),\\
-\dot{p_2} &=& \dot{\nu}(\rho_2+p_2),\\
e^{2\mu}\rho_1 &=& 2\dot{\lambda}\dot{\mu}
-3{\dot{\lambda}}^{2}-2\ddot{\lambda} ,\\
e^{2\nu}\rho_2 &=& 2\dot{\lambda}\dot{\mu}
+{\dot{\lambda}}^{2},\\
0&=&\ddot{\lambda}+{\dot{\lambda}}^2
-\dot{\lambda}\dot{\nu}-\dot{\lambda}\dot{\mu},\\
e^{2\mu}p_1&=&{\dot{\lambda}}^2+2\dot{\lambda}\dot{\nu},\\
e^{2\nu}p_2&=& -2\ddot{\lambda}-3{\dot{\lambda}}^2
-2\dot{\lambda}+2\dot{\lambda}\dot{\nu},\\
e^{2\mu}p_1&=&\ddot{\lambda}+{\dot{\lambda}}^2+\dot{\lambda}\dot{\nu}
+\ddot{\nu}+{\dot{\nu}}^2+\dot{\lambda}\dot{\mu}-\dot{\mu}\dot{\nu},\\
e^{2\nu}p_2&=&-\ddot{\lambda}-{\dot{\lambda}}^2
-\dot{\lambda}+\dot{\lambda}\dot{\nu}-\ddot{\mu}-{\dot{\mu}}^2
-\dot{\mu}+\dot{\mu}\dot{\nu}-\dot{\lambda}\dot{\mu}.
\end{eqnarray}

\subsubsection{EOS(1) and EOS(2)}

For EOS(1), Eqs.(81) and (82) imply that
\begin{eqnarray}
p_2=0=\rho_2,\quad p_1= k(8\pi
G)^{(1-\gamma)}{\rho_1}^\gamma.\quad [Case~I]
\end{eqnarray}
EOS(2) implies that
\begin{eqnarray}
p_2=0=\rho_2,\quad p_1=\frac{k}{{m_b}^{\gamma}(8\pi
G)^{(\gamma-1)}}{(\rho_1-\frac{p_1}{(\gamma-1)})}^\gamma.\quad
[Case~II]
\end{eqnarray}
In both cases, Eq.(88) shows that either $\lambda=constant$ or
$\dot{\lambda}=-2\dot{\mu}$. If $\lambda=constant$, Eqs.(86) and
(90), respectively, imply that $\rho_1=0=p_1$ and we are left with
Eqs.(92) and (93). Solving these two equations lead to
$\ddot{\nu}+{\dot{\nu}}^2-\ddot{\mu}-{\dot{\mu}}^2-\dot{\mu}=0$
which satisfies for four different possibilities. For
$\dot{\mu}=0=\dot{\nu}$, we trivially get \textit{Minkowski}
spacetime. For $\dot{\mu}=0$, the solution turns out to be
\begin{eqnarray}
\nu&=&\ln(\ln{\xi}-\ln{c_1})+c_2,\quad \mu=c_3,
\quad  \lambda=c_4,\nonumber\\
\rho_1&=&0=p_1,\quad \rho_2=0=p_2.
\end{eqnarray}
The metric will be
\begin{equation}
ds^2=[\ln{(\frac{x}{c_1t})}]^2dt^2-\frac{1}{x^2}dx^2-(dy^2+
dz^2),\quad(c_1\neq0).
\end{equation}
In the case $\dot{\nu}=0$, we obtain
\begin{eqnarray}
\nu&=&c_1,\quad \mu=\ln(\ln{\xi}-\ln{c_2})+c_3,
\quad  \lambda=c_4,\nonumber\\
\rho_1&=&0=p_1,\quad \rho_2=0=p_2.
\end{eqnarray}
The corresponding metric is
\begin{equation}
ds^2=dt^2-\frac{1}{x^2}[\ln{(\frac{x}{c_2t})}]^2dx^2-(dy^2+
dz^2),\quad(c_2\neq0).
\end{equation}
Finally, for the last possibility $\ddot{\nu}+{\dot{\nu}}^2=0$ and
$-\ddot{\mu} -{\dot{\mu}}^2-\dot{\mu}=0$, Eqs.(92) and (93) imply
$\dot{\mu}\dot{\nu}=0$ and again we have the above possibilities.
The second case, when $\dot{\lambda}=-2\dot{\mu}$, gives the same
solution as given by Eq.(97).

\subsubsection{EOS(3)}

Eqs.(81) and (82) imply that
\begin{equation}
p_1=k\rho_1,\quad  p_2=k\rho_2.\quad [Case~III]
\end{equation}
When $k=-1$, this gives rise to the same solution as for EOS(1) and
EOS(2). The second case, i.e., $k\neq-1$ also leads to the same
results as in EOS(1) and EOS(2).

\section{Tilted Dust Case}

\subsection{Self-Similarity of the First Kind}

When we take $p=0$ in Eqs.(22)-(30) for the tilted perfect fluid
case with self-similarity of the first kind, Eq.(23) gives either
$\dot{\nu}=0$ or $\rho=0$. Both the cases yield contradiction.

\subsection{Self-Similarity of the Second Kind}

Here for $p_1=0=p_2$, Eqs.(38) and (39) imply that either
$\nu=constant$ or $\rho_1=0=\rho_2$. For the first possibility, we
obtain the following solution
\begin{eqnarray}
\nu&=&c_1,\quad
\mu=\ln{(c_3\xi^{-1/2}(\xi^{3/2}\mp2{c_2}^{3/2}))}, \quad
\lambda=-\ln{\xi}+c_4,\nonumber\\ \rho_1&=&0=p_1,\quad
\rho_2=\frac{2}{3c_5}(2-3(\frac{\xi^{3/2}}{\xi^{3/2}\mp2{c_2}^{3/2}})),\quad
p_2=0, \nonumber\\\alpha &=&\frac{3}{2}.
\end{eqnarray}
The corresponding metric is
\begin{equation}
ds^2=dt^2-(\frac{3t^{2/3}}{2^{2/3}x})(\frac{2x^{3/2}}{3t}
\mp2{c_2}^{3/2})^{2}dx^2-(\frac{3t}{2})^{4/3}(dy^2+dz^2).
\end{equation}
The second possibility leads to the same solution as given by Eq.(51).

\subsection{Self-Similarity of the Zeroth Kind}

This case gives contradiction and hence there is no solution.

\subsection{Self-Similarity of the Infinite Kind}

In this case, Eqs.(85) and (86) imply that either $\nu=constant$ or
$\rho_1=0=\rho_2$. In the first case, we obtain the following
solution
\begin{eqnarray}
\nu&=&c_1,\quad \mu=-\ln{\xi}+\ln(\xi-c_2)+c_3,
\quad  \lambda=c_4,\nonumber\\
\rho_1&=&0=\rho_2.
\end{eqnarray}
The corresponding metric is
\begin{equation}
ds^2=dt^2-\frac{(x-c_2t)^2}{x^4}dx^2-(dy^2+ dz^2).
\end{equation}
For the second case, when $\rho_1=0=\rho_2$, Eqs.(88) and (90) imply
that either $\dot{\lambda}=0$ or
$\dot{\mu}=\dot{\nu},~\dot{\lambda}=-2\dot{\mu}$. The first option
yields exactly the same result as for the tilted perfect fluid with
self-similarity of the infinite kind using EOS(1) and EOS(2) and are
given by Eqs.(96), (98) and \textit{Minkowski} spacetime. The other
possibility implies a \textit{Minkowski} spacetime.

\section{Orthogonal Perfect Fluid and Dust Cases}

Here the self-similar variable is $\xi=t$ in each kind. The EFEs and
the equations of motion for the perfect fluid of the first kind
gives the following set of equations
\begin{eqnarray}
\dot{\mu}&=&0,\\
e^{2\nu}(e^{-2\mu}+\rho)&=& {\lambda'}^2,\\
e^{2\nu}(3e^{-2\mu}-p)&=&3{\lambda'}^2+2\lambda''-2\lambda'\nu',\\
e^{2\nu}(e^{-2\mu}-p)&=&\lambda''+{\lambda'}^2-\lambda'\nu',\\
2\lambda'(\rho+p)&=&-\rho'_1,\\
\rho&=&p,
\end{eqnarray}
where prime indicates derivative with respect to $\xi=t$. Eq.(110)
gives an equation of state for this system of equations. Solving
these equations simultaneously, we arrive at the following solution
\begin{eqnarray}
\nu&=&\ln{(\frac{p'}{4p\sqrt{(c_0+p)}})},\quad \mu=c_1,\quad
\lambda=-\frac{1}{4}\ln{(p)}+\ln{(c_2)},\nonumber\\
\rho&=&p,\quad {p'}^2p-2(1+p)(p''p-{p'}^2)=0.
\end{eqnarray}
For the perfect fluid case of the second and zeroth kinds, we obtain
contradiction. The perfect fluid case of the infinite kind gives
\emph{Minkowski} spacetime.

For the dust case, we take $p=0$ in the equations for the perfect
fluid case. In the self-similarity of the first kind, Eq.(110) shows
that the resulting spacetime must be vacuum. Eq.(106) gives
$e^{2\nu}e^{2\mu}={\lambda'}^2$ and we obtain
\begin{eqnarray}
\nu=\nu(\xi), \quad \lambda=c_0\int e^{\nu(\xi)}d\xi,\quad \mu= c_1,
\quad \rho=0=p.
\end{eqnarray}
The metric becomes
\begin{equation}
ds^2=x^2e^{2\nu(t)}dt^2-dx^2-x^2e^{2c_0\int e^{\nu(t)}dt}(dy^2+
dz^2).
\end{equation}
For the self-similarity of the second, zeroth and infinite kinds, we
arrive at a contradiction due to one or the other reason and hence
there is no solution.

\section{Parallel Perfect Fluid Case}

\subsection{Self-Similarity of the First Kind}

Here the self-similar variable is $\xi=x$ and the metric functions
are given by Eq.(17). A set of ODEs in terms of $\xi$ are
\begin{eqnarray}
\nu'&=&0,\\
\rho&=& 3e^{-2\nu}+e^{-2\mu}(2\lambda'\mu'-3{\lambda'}^2-2\lambda''),\\
p &=&e^{-2\mu}({\lambda'}^2+2\lambda'\nu')-e^{-2\nu},\\
p&=&e^{-2\mu}(\lambda''+{\lambda'}^2+\lambda'\nu'+\nu''+{\nu'}^2
-\lambda'\mu'-\mu'\nu')-e^{-2\nu},\\
0&=&\rho+3p,\\
-p'&=&\nu'(\rho+p).
\end{eqnarray}
Here prime denotes derivative with respect to $\xi=x$. Eq.(118)
indicates an equation of state. Using Eq.(114) in rest of the
equations, we get $p'=0$. Solving the remaining equations, we obtain
\begin{eqnarray}
\nu=c_1, \quad \mu= c_2, \quad \lambda=c_3\xi+c_4,\quad \rho=0=p
\end{eqnarray}
and the corresponding spacetime is
\begin{equation}
ds^2=dt^2-t^2dx^2-t^2e^{2c_3x}(dy^2+ dz^2).
\end{equation}

\subsection{Self-Similarity of the Second Kind}

For this kind, the self-similar variable is also $\xi=x$ and the
metric functions are given by Eq.(17). The EFEs imply that the
quantities $\rho$ and $p$ must be of the form
\begin{eqnarray}
\kappa\rho &=&t^{-2}\rho_1(\xi)+t^{-2\alpha}\rho_2(\xi),\\
\kappa p &=& t^{-2}p_1(\xi)+t^{-2\alpha}p_2(\xi).
\end{eqnarray}
A set of ODEs in terms of $\xi$ will be
\begin{eqnarray}
\nu'&=&0,\\
e^{2\mu}\rho_1&=& 2\lambda'\mu'-3{\lambda'}^2-2\lambda'',\\
\rho_2&=&3e^{-2\nu},\\
e^{2\mu}p_1 &=& {\lambda'}^2,\\
e^{2\nu}p_2 &=&2\alpha-3,\\
e^{2\mu}p_1&=& \lambda''+{\lambda'}^2-\lambda'\mu',\\
e^{2\nu}p_2&=&2\alpha-3,\\
0&=&\rho_1+3p_1,\\
0&=&(3-2\alpha)\rho_2+3p_2,\\
-p'_1&=&0,\\
-p'_2&=&0.
\end{eqnarray}

\subsubsection{EOS(1) and EOS(2)}

When a perfect fluid satisfies EOS(1), Eqs.(122) and (123) imply
that
\begin{eqnarray}
p_2=0=\rho_1,\quad \alpha=\frac{1}{\gamma},\quad p_1= k(8\pi
G)^{(1-\gamma)}{\rho_2}^\gamma.\quad [Case~I]
\end{eqnarray}
For EOS(2), it turns out that
\begin{eqnarray}
p_2=0,\quad \alpha=\frac{1}{\gamma},\quad
p_1=\frac{k}{{m_b}^{\gamma}(8\pi
G)^{(\gamma-1)}}{\rho_2}^\gamma=(\gamma-1)\rho_1.\quad [Case~II]
\end{eqnarray}
The \textbf{Case $I$} gives a contradiction and the \textbf{Case
$II$} yields the following solution
\begin{eqnarray}
\nu&=&c_1,\quad \mu=\ln{\lambda'}+c_2, \quad \lambda=\lambda(\xi),\nonumber\\
\rho_1&=&-3p_1=constant, \quad \rho_2=\frac{3}{{c_0}^2}, \quad
p_2=0, \quad \alpha=\frac{3}{2}.
\end{eqnarray}
The spacetime becomes
\begin{equation}
ds^2=tdt^2-t^2{\lambda'}^2dx^2-t^2e^{2\lambda(x)}(dy^2+ dz^2).
\end{equation}

\subsubsection{EOS(3)}

For EOS(3), Eqs.(122) and (123) show that
\begin{equation}
p_1=k\rho_1,\quad  p_2=k\rho_2.
\end{equation}
Here Eqs.(131) and Eq.(132) imply that $\rho_1=0$ and Eq.(124)
gives $\dot{\nu}=0$ while Eq.(127) implies that $\dot{\lambda}=0$.
Solving the remaining equations, it turns out that
\begin{eqnarray}
\nu&=&c_1,\quad \mu=\mu(\xi),
\quad  \lambda=c_2,\quad \rho_1=0=p_1,\nonumber\\
\quad \rho_2&=&\frac{3}{c_0}, \quad p_2=-\frac{(3-2\alpha)}{c_0},
\quad k=-\frac{(3-2\alpha)}{3}.
\end{eqnarray}
This gives the following spacetime
\begin{equation}
ds^2=t^{2(\alpha-1)}dt^2-t^2e^{2\mu(x)}dx^2-t^2(dy^2+ dz^2).
\end{equation}

\subsection{Self-Similarity of the Zeroth Kind}

For this kind, the self-similar variable is again $\xi=x$ and the
plane symmetric metric functions are given by Eq.(17). The EFEs
imply that the quantities $\rho$ and $p$ must be of the form
\begin{eqnarray}
\kappa\rho &=&t^{-2}\rho_1(\xi)+\rho_2(\xi),\\
\kappa p &=& t^{-2}p_1(\xi)+p_2(\xi).
\end{eqnarray}
ODEs are
\begin{eqnarray}
\nu'&=&0,\\
e^{2\mu}\rho_1&=& 2\lambda'\mu'-3{\lambda'}^2-2\lambda'',\\
\rho_2&=&3e^{-2\nu},\\
e^{2\mu}p_1 &=& {\lambda'}^2,\\
e^{2\nu}p_2 &=&-3,\\
e^{2\mu}p_1&=& \lambda''+{\lambda'}^2-\lambda'\mu',\\
e^{2\nu}p_2&=&-3,\\
0&=&\rho_1+3p_1,\\
0&=&\rho_2+p_2,\\
-p'_1&=&0,\\
-p'_2&=&0.
\end{eqnarray}

\subsubsection{EOS(1) and EOS(2)}

In the case of EOS(1), Eqs.(142) and (143) yield
\begin{eqnarray}
\rho_1=0=p_1,\quad  p_2= k(8\pi G)^{(1-\gamma)}{\rho_2}^\gamma.\quad [Case~I]
\end{eqnarray}
For EOS(2), it turns out that
\begin{eqnarray}
p_1=0=\rho_1,\quad p_2=\frac{k}{{m_b}^{\gamma}(8\pi
G)^{(\gamma-1)}}[{\rho_2}-\frac{p_2}{(\gamma-1)}]^\gamma,
\quad [Case~II]
\end{eqnarray}
In both the cases, we obtain the same solution as
\begin{eqnarray}
\nu&=&c_1,\quad \mu=\mu(\xi), \quad \lambda=c_2,\nonumber\\
\rho_1&=&0=p_1, \quad \rho_2=-p_2=constant
\end{eqnarray}
and the resulting metric is
\begin{equation}
ds^2=\frac{1}{t^2}dt^2-t^2e^{2\mu(x)}dx^2-t^2(dy^2+ dz^2).
\end{equation}

\subsubsection{EOS(3)}

Eqs.(142) and (143) show that
\begin{equation}
p_1=k\rho_1,\quad  p_2=k\rho_2.
\end{equation}
Here Eqs.(151) and (152) imply that either $\rho_1=0$ or
$\rho_2=0$. Eq.(146) gives a contradiction for $\rho_2=0$ and
hence $\rho_1=0$. Also, Eq.(151) shows that $k=-1$ hence this
gives the same solution as in EOS(1) and EOS(2) given by Eq.(158).

\subsection{Self-Similarity of the Infinite Kind}

Again we have the self-similar variable $\xi=x$ and the spacetime
metric coefficients are given by Eq.(17). A set of ODEs will be
\begin{eqnarray}
-e^{2\mu}\rho&=&
3{\lambda'}^2+2\lambda''-2\lambda'\mu',\\
e^{2\mu}p &=&{\lambda'}^2+2\lambda'\nu',\\
e^{2\mu}p&=&\lambda''+{\lambda'}^2+\lambda'\nu'+\nu''
+{\nu'}^2-\lambda'\mu'-\nu'\mu',\\
-p'&=&\nu'(\rho+p).
\end{eqnarray}
We consider the following four possibilities to solve the above
set of equations.
\begin{eqnarray*}
&(i)& \quad  \nu'=\mu', \quad (ii) \quad \nu'=\lambda',\\
&(iii)& \quad \lambda'=\mu', \quad (iv) \quad \nu'=\lambda'=\mu'.
\end{eqnarray*}
The first case gives the following solution
\begin{eqnarray}
\nu&=&\mu=c_1,\quad \lambda=c_2\xi+c_3,\nonumber\\
\rho&=&-3p=constant
\end{eqnarray}
and the corresponding metric is
\begin{equation}
ds^2=dt^2-dx^2-e^{2x}(dy^2+ dz^2).
\end{equation}
The second case corresponds to \emph{Minkowski} spacetime. For the
case (iii), we obtain the following solution
\begin{eqnarray}
\nu&=&c_1,\quad \lambda=\mu=c_3-\ln{(\xi-c_2)},\nonumber\\
\rho&=&3p, \quad p=-1
\end{eqnarray}
and the metric is given by
\begin{equation}
ds^2=dt^2-\frac{1}{x^2}(dx^2+dy^2+ dz^2).
\end{equation}
The last case yields \emph{Minkowski} spacetime.

\section{Parallel Dust case}

\subsection{Self-Similarity of the First Kind}

Setting $p=0$ in the equations for the parallel perfect fluid case
with self-similarity of the first kind, we finally have a
contradiction and hence we do not have any self-similar solution.

\subsection{Self-Similarity of the Second Kind}

For $p_1=0=p_2$, Eqs.(124) and (134) show that
$\nu=constant=\lambda$ respectively and we get the same solution
as given by Eq.(157) with $\rho_2=0=p_2$ and $\alpha=\frac{3}{2}$
but the corresponding metric is
\begin{equation}
ds^2=tdt^2-t^2e^{2\mu(x)}dx^2-t^2(dy^2+ dz^2).
\end{equation}

\subsection{Self-Similarity of the Zeroth Kind}

When we take $p_1=0=p_2$, Eqs.(148) and (150) lead to
contradiction.

\subsection{Self-Similarity of the Infinite Kind}

For $p=0$, Eq.(163) shows that either $\nu=constant$ or $\rho=0$. In
the first case, the resulting spacetime is \textit{Minkowski}. For
$\rho=0$, Eq.(161) implies that either $\lambda'=0$ or
$\lambda'=-2\nu'$. When $\lambda'=0$, we obtain
$\nu''+{\nu'}^2-\mu'\nu'=0$ which implies that either $\nu'=0$ or
$\mu'=0$. For the first possibility, we obtain \textit{Minkowski}
spacetime. For the second possibility, we get the following vacuum
solution
\begin{eqnarray}
\nu=ln(c_2(\xi-c_1)),\quad \mu=c_3,\quad\lambda=c_4,\quad
\rho=0=p.
\end{eqnarray}
The metric for this spacetime is
\begin{equation}
ds^2=(c_2(x-c_1))^2dt^2-(dx^2+dy^2+ dz^2).
\end{equation}
For $\lambda'=-2\nu'$, Eqs.(160) and (162) imply that
$2\nu''-3{\nu'}^2-\mu'\nu'=0$ which gives either $\nu'=0$ or
$\mu'=0$. The first possibility leads to the \textit{Minkowski}
spacetime and the second possibility gives the following vacuum
solution
\begin{eqnarray}
\nu=ln(\frac{c_2}{(3x-c_1)^{1/3}}),\quad \mu=c_3,\quad\lambda=c_4,
\quad \rho=0=p
\end{eqnarray}
and the corresponding metric is
\begin{equation}
ds^2=(\frac{{c_2}^2}{(3x-c_1)^{2/3}})dt^2-(dx^2+dy^2+ dz^2).
\end{equation}

\section{Summary and Discussion}

Recent literature [3-5,7,17-20] indicates keen interest in the
self-similar solutions and their physical features. Maeda et al.
[3-5] have classified the spherically symmetric KSS perfect fluid
and dust solutions. Sharif and Sehar [7] have extended this analysis
for the classification of the KSS cylindrically symmetric solutions.
Recently, the same authors [21] have explored the KSS solutions for
the plane symmetric spacetimes under certain restriction, i.e.,
$\mu=0$ for the sake of simplicity. Consequently, the classification
was incomplete in the sense that we were missing many such cases
where solution can be possible. This paper deals with the most
general plane symmetric spacetimes and provides self-similar
solutions even in those cases where we obtain null results [21]. We
have classified KSS perfect fluid and dust solutions for the cases
when KSS vector is tilted, orthogonal and parallel to the fluid flow
by using EOS(1), EOS(2) and EOS(3). This gives rise to twenty four
plane symmetric self-similar solutions out of which we obtain
sixteen independent solutions.

It is found that EOS(1) and EOS(2) are incompatible with the
self-similarity of the first kind in the tilted perfect fluid case.
For EOS(3), we obtain solution with constant density. For the
self-similarity of the second kind with EOS(1) and EOS(2), we obtain
a vacuum solution. For EOS(3) with $k=-1$, it follows the same
solution as for EOS(1) and EOS(2). The case $k\neq-1$ leads to two
self-similar solutions one of these ($\rho_1=0$) represents a stiff
fluid. The zeroth kind with EOS(1) and EOS(2) yields a solution.
EOS(3) gives three solution, one is vacuum solution and other is a
stiff fluid solution. In the case of the infinite kind for EOS(1)
and EOS(2), we find three vacuum solutions while EOS(3) also leads
to vacuum solutions both for $k=-1$ and $k\neq-1$.

For the tilted dust case with self-similarity of the second kind,
we obtain the same solution as for the tilted perfect fluid with
EOS(1) and EOS(2) and a dust solution for $\alpha=3/2$. The
self-similarity of the infinite kind leads to four vacuum
solutions one of them is \textit{Minkowski} spacetime. There is no
solution in any other kind.

In the orthogonal perfect fluid with self-similarity of the first
kind we obtain a solution in terms of pressure and with
self-similarity of the infinite kind we obtain \textit{Minkowski}
spacetime. Any other kind does not provide any solution. The
orthogonal dust case with self-similarity of the first kind yields a
vacuum spacetime given by Eq.(112). All other kinds provide
contradictory results.

In the parallel perfect fluid with the self-similarity of the first
kind, we obtain a vacuum solution. The second kind leads to
contradiction for EOS(1) but for EOS(2), we obtain a solution in
which one fluid represents dust and the other vacuum. For EOS(3), we
obtain a solution with arbitrary $\mu$ and $\alpha\neq\frac{3}{2}$.
The zeroth kind yields a vacuum solution. There are three
self-similar solution with self-similarity of the infinite kind one
of which is \textit{Minkowski} spacetime. For the parallel dust
case, the first kind gives a contradiction. The second kind implies
the same solution as in the parallel perfect fluid case with EOS(3)
and $p_2=0,~\alpha=\frac{3}{2}$. We do not have any self-similar
solution of zeroth kind in the dust case. However, we obtain three
different vacuum solutions for the infinite kind.

It is interesting to note that all the self-similar solutions,
except the solutions given by Eqs.(33), (80), (97), (99), (104),
(111), (165), (167), (170), (172) found here correspond with the
already classified solutions [23] under particular coordinate
transformations. The metrics given by Eqs.(52), (55), (74), (79) and
(102) correspond to the class of metrics
\begin{equation}
ds^2=dt^2-e^{2\mu(t)}dx^2-e^{2\lambda(t)}(dy^2+ dz^2).
\end{equation}
This metric has the four KVs admitting $G_3\otimes\Re$ with a
spacelike $\Re$ and can be matched with Kantowski Sachs spacetimes
[22]. The spacetime given by Eq.(57) correspond to the class of
metrics
\begin{equation}
ds^2=e^{2\nu(x)}dt^2-dx^2-e^{2\lambda(x)}(dy^2+ dz^2)
\end{equation}
which has four KVs with the same Lie algebra and a timelike $\Re$.
The solution given by Eq.(113) can be matched with the solution
\begin{equation}
ds^2=e^{2f(x)}[dt^2-e^{2t/a}(dy^2+ dz^2)]-dx^2,\quad (a\neq0)
\end{equation}
which admits six KVs. The metrics (121) and (138) turn out to be
equivalent to the metric
\begin{equation}
ds^2=dt^2-e^{2f(t)}[dx^2+e^{2x/a}(dy^2+ dz^2)],\quad (a\neq0)
\end{equation}
which has six KVs with a Lie algebra identical to that of the metric
(175) and belongs to the family of LRS metrics. Finally, the metrics
given by Eqs.(77), (141), (158) and (168) has the correspondence
with the class of metrics given as
\begin{equation}
ds^2=dt^2-e^{\lambda(t)}(dx^2+dy^2+ dz^2)
\end{equation}
admitting six KVs and represents FRW models. We also notice that the
solutions given by the metrics (99), (104) seem to have the similar
nature while the solutions (165) and (167), and solutions (170) and
(172) can correspond to each other. It is worth mentioning that we
obtain density either zero or positive in all the solutions except
the one where it is not constant but can be positive. The physical
properties of such solutions can be seen in [20]. Thus we finally
obtain sixteen independent KSS plane symmetric solutions. The
results can be summarized in the form of tables 1-6:

\vspace{0.5cm}

{\bf {\small Table 1.}} {\small Tilted Perfect Fluid KSS Solutions.}

\vspace{0.5cm}

\begin{center}
\begin{tabular}{|l|l|}
\hline {\bf Self-Similarity} & {\bf Solution}
\\ \hline First kind (EOS(3)) & solution given by Eq.(33)
\\ \hline Second kind (EOS(1)) & solution given by Eq.(52)
\\ \hline Second kind (EOS(2)) & solution given by Eq.(52)
\\ \hline Second kind (EOS(3))(i) & solution given by Eq.(52)
\\ \hline Second kind (EOS(3))(ii) & solution given by Eq.(55)
\\ \hline Second kind (EOS(3))(iii) & solution given by Eq.(57)
\\ \hline Zeroth kind (EOS(1)) & solution given by Eq.(74)
\\ \hline Zeroth kind (EOS(2)) & solution given by Eq.(74)
\\ \hline Zeroth kind (EOS(3))(i) & solution given by Eq.(77)
\\ \hline Zeroth kind (EOS(3))(ii) & solution given by Eq.(79)
\\ \hline Zeroth kind (EOS(3))(iii) & solution given by Eq.(80)
\\ \hline Infinite kind (EOS(1))(i) & Minkowski spacetime
\\ \hline Infinite kind (EOS(1))(ii) & solution given by Eq.(97)
\\ \hline Infinite kind (EOS(1))(iii) & solution given by Eq.(99)
\\ \hline Infinite kind (EOS(2)) & Same solutions as in EOS(1)
\\ \hline Infinite kind (EOS(3)) & Same solutions as in EOS(1)
\\ \hline
\end{tabular}
\end{center}

\newpage

{\bf {\small Table 2.}} {\small Tilted Dust KSS Solutions.}

\vspace{0.5cm}

\begin{center}
\begin{tabular}{|l|l|}
\hline {\bf Self-similarity} & {\bf Solution}
\\ \hline First kind  & None
\\ \hline Second kind(i)& solution given by Eq.(102)
\\ \hline Second kind(ii)& solution given by Eq.(52)
\\ \hline Zeroth kind  & None
\\ \hline Infinite kind  (i)& solution given by Eq.(104)
\\ \hline Infinite kind  (ii)& Minkowski spacetime
\\ \hline Infinite kind  (iii)& solution given by Eq.(97)
\\ \hline Infinite kind  (iv)& solution given by Eq.(99)
\\ \hline
\end{tabular}
\end{center}

\vspace{0.5cm}

{\bf {\small Table 3.}} {\small Orthogonal Perfect Fluid KSS
Solutions.}

\vspace{0.5cm}

\begin{center}
\begin{tabular}{|l|l|}
\hline {\bf Self-Similarity} & {\bf Solution}
\\ \hline First kind & solution given by Eq.(111)
\\ \hline Second kind & None
\\ \hline Zeroth kind & None
\\ \hline Infinite kind & Minkowski spacetime
\\ \hline
\end{tabular}
\end{center}

\vspace{0.5cm}

{\bf {\small Table 4.}} {\small Orthogonal Dust KSS Solutions.}

\vspace{0.5cm}

\begin{center}
\begin{tabular}{|l|l|}
\hline {\bf Self-similarity} & {\bf Solution}
\\ \hline First kind  & solution given by Eq.(113)
\\ \hline Second kind & None
\\ \hline Zeroth kind  & None
\\ \hline Infinite kind & None
\\ \hline
\end{tabular}
\end{center}

\newpage
{\bf {\small Table 5.}} {\small Parallel Perfect Fluid KSS
Solutions.}

\vspace{0.5cm}

\begin{center}
\begin{tabular}{|l|l|}
\hline {\bf Self-similarity} & {\bf Solution}
\\ \hline First kind  & solution given by Eq.(121)
\\ \hline Second kind (EOS(1)) & None
\\ \hline Second kind (EOS(2)) & solution given by Eq.(138)
\\ \hline Second kind (EOS(3)) & solution given by Eq.(141)
\\ \hline Zeroth kind (EOS(1)) & solution given by Eq.(158)
\\ \hline Zeroth kind (EOS(2)) & solution given by Eq.(158)
\\ \hline Zeroth kind (EOS(3)) & solution given by Eq.(158)
\\ \hline Infinite kind (i) & solution given by Eq.(165)
\\ \hline Infinite kind (ii) & Minkowski spacetime
\\ \hline Infinite kind (iii)& solution given by Eq.(167)
\\ \hline Infinite kind (iv) & Minkowski spacetime
\\ \hline
\end{tabular}
\end{center}

\vspace{0.5cm}

{\bf {\small Table 6.}} {\small Parallel Dust KSS Solutions.}

\vspace{0.5cm}

\begin{center}
\begin{tabular}{|l|l|}
\hline {\bf Self-similarity} & {\bf Solution}
\\ \hline First kind  & None
\\ \hline Second kind  & solution given by Eq.(168)
\\ \hline Zeroth kind  & None
\\ \hline Infinite kind (i) & Minkowski spacetime
\\ \hline Infinite kind (ii) & Minkowski spacetime
\\ \hline Infinite kind (iii) & solution given by Eq.(170)
\\ \hline Infinite kind (iv) & Minkowski spacetime
\\ \hline Infinite kind (v) & solution given by Eq.(172)
\\ \hline
\end{tabular}
\end{center}

Finally, we would like to mention that Sintes et al. [14] found
solutions only for the infinite kind. However, we have studied KSS
solutions of the most general plane symmetric spacetimes in all
kinds. The KSS solutions of the infinite kind can be matched with
those of [14]. The solutions given by Eqs.(99) and (104) can be
matched with Eq.(5.5) and the solutions (97), (170) and (172)
correspond to the solution (5.6) of [14]. The remaining solutions do
not correspond to those of the solutions given in [14].

\begin{description}
\item  {\bf Acknowledgment}
\end{description}

One of us (SA) would like to thank HEC for Merit Scholarship.

\newpage

{\bf \large References}

\begin{description}

\item{[1]} Carr, B.J.: Phys. Rev. {\bf D62}(2000)044022.

\item{[2]} Carr, B.J. and Coley, A.A.: Phys. Rev. {\bf D62}(2000)044023.

\item{[3]} Maeda, H., Harada, T., Iguchi, H. and Okuyama, N.: Phys. Rev.
           {\bf D66}(2002)027501.

\item{[4]} Maeda, H., Harada, T., Iguchi, H. and Okuyama, N.: Prog. Theor. Phys.
           {\bf108}(2002)819.

\item{[5]} Maeda, H., Harada, T., Iguchi, H. and Okuyama, N.: Prog. Theor. Phys.
           {\bf110}(2003)25.

\item{[6]} Sharif, M.: J. Math. Phys. {\bf44}(2003)5141;
ibid {\bf45}(2004)1518; ibid {\bf45}(2004)1532.

\item{[7]} Sharif, M. and Aziz, Sehar: Int. J. Mod. Phys. {\bf D14}(2005)1527;\\
           {\it Kinematic Self-Similar Solutions: Proceedings of the 11th
           Regional Conference on Mathematical
           Physics and IPM Spring Conference} Tehran-Iran, May, 3-6, 2004,
           eds. Rahvar, S., Sadooghi, N. and Shojai, F. (World Scientific,
           2005)111.

\item{[8]} Cahill, M.E. and Taub, A.H.: Commun. Math. Phys.
           {\bf21}(1971)1.

\item{[9]} Carter, B. and Henriksen, R.N.: Annales De Physique
           {\bf14}(1989)47.

\item{[10]} Carter, B. and Henriksen, R.N.: J. Math. Phys.
           {\bf32}(1991)2580.

\item{[11]} Coley, A.A.: Class. Quant. Grav. {\bf14}(1997)87.

\item{[12]} McIntosh, C.B.G.: Gen. Relat. Gravit. {\bf7}(1975)199.

\item{[13]} Benoit, P.M. and Coley, A.A.: Class. Quant. Grav.
           {\bf15}(1998)2397.

\item{[14]} Sintes, A.M., Benoit, P.M. and Coley, A.A.: Gen. Relat. Gravit.
           {\bf33}(2001)1863.

\item{[15]} Carr, B.J., Coley, A.A., Golaith, M., Nilsson, U.S. and Uggla, C.:
           Class. Quant. Grav. {\bf18}(2001)303.

\item{[16]} Carr, B.J., Coley, A.A., Golaith, M., Nilsson, U.S. and Uggla, C.:
            Phys. Rev. {\bf D61}(2000)081502.

\item{[17]} Coley, A.A. and Golaith, M.: Class. Quant. Grav. {\bf17}(2000)2557.

\item{[18]} Sharif, M. and Aziz, Sehar: Int. J. Mod. Phys. {\bf D14}(2005)73.

\item{[19]} Sharif, M. and Aziz, Sehar: Int. J. Mod. Phys. {\bf A20}(2005)7579.

\item{[20]} Sharif, M. and Aziz, Sehar : J. Korean Physical Society {\bf 47}(2005)757.

\item{[21]} Sharif, M. and Aziz, Sehar: J. Korean Physical Society {\bf 49}(2006)21.

\item{[22]} Stephani, H., Kramer, D., Maccallum, M., Hoenselaers, C. and Herlt, E.
            \textit{Exact Solutions of Einstein's Field Equations}
            (Cambridge University Press, 2003).
\item{[23]} Feroze, Tooba, Qadir, A. and Ziad, M.: J. Math. Phys. {\bf 42}(2001)4947.

\end{description}

\end{document}